\pdfoutput=1

\documentclass[10pt,fleqn]{article}

\usepackage{latexsym}
\usepackage{alltt}
\usepackage{tabularx}
\usepackage{color}

\usepackage[margin=2cm]{geometry}

\usepackage{lmodern}            
\usepackage{enumitem}

\newcommand{\cfr}{\emph{cfr.}}

\newcommand{\CL}{\textsf{Common~Lisp}}

\newcommand{\CMUCL}{\textsf{CMUCL}}
\newcommand{\SBCL}{\textsf{SBCL}}

\newcommand{\CLang}{\textsf{C}}
\newcommand{\Cpp}{\textsf{C++}}
\newcommand{\Java}{\textsf{\textsc{Java}}}
\newcommand{\Fortran}{\textsc{\textsf{Fortran}}}
\newcommand{\RLang}{\textsf{R}}
\newcommand{\Python}{\textsc{\textsf{Python}}}

\newcommand{\code}[1]{\texttt{#1}}
\newcommand{\codeit}[1]{\textit{\texttt{#1}}}

\newcommand{\clieeeterm}[1]{\textit{#1}}

\newcommand{\varname}[1]{\textit{#1}}

\newcommand{\clterm}[1]{\textit{#1}}

\newcommand{\codeprompt}[1]{\textcolor{blue}{\textbf{#1}}}
\newcommand{\codelia}[1]{\textcolor{blue}{#1}}

\newcommand{\RArrow}{$\Rightarrow$}

\newcommand{\IEEEFPStd}{IEEE-754}
\newcommand{\IECFPStd}{IEC-60559}

\newcommand{\DDictionaryItem}[1]{\vspace*{6pt}\noindent\hrulefill\vspace*{-9pt}\subsection*{#1}}

\newcommand{\DSyntax}{\subsubsection*{Syntax:\\}}

\newcommand{\DArgsNValues}{\subsubsection*{Arguments and Values:\\}}
\newcommand{\DDescription}{\subsubsection*{Description:\\}}

\newcommand{\DExceptional}{\subsubsection*{Exceptional Situations:\\}}

\title{\LARGE{\bfseries Why You Cannot (Yet) Write an ``Interval Arithmetic''
  Library in \CL{}}\\\ldots or: Hammering Some Sense into
\code{:ieee-floating-point}}

\author{
  Marco Antoniotti\\
  Dipartimento di Informatica, Sistemistica e Comunicazione\\
  Universit\`{a} degli Studi di Milano Bicocca\\
  Viale Sarca 336, U14, Milan (MI), \textsc{Italy}\\[2mm]
  \texttt{marco.antoniotti} at \texttt{unimib.it},\\
  \texttt{mantoniotti} at \texttt{common-lisp.net}}

\begin{document}

\maketitle

\begin{abstract}
  ``Interval Arithmetic'' (IA) appears to be a useful numerical tool
  to have at hand in several applications.  Alas, the current IA
  descriptions and proposed standards are always formulated in terms
  of the \IEEEFPStd{} standard, and the status of \IEEEFPStd{}
  compliance of most \CL{} implementations is not up to par.

  A solution would be for \CL{} implementations to adhere to the
  \emph{Language Independent Arithmetic} (LIA) IEC standard, which
  includes IEEE 754.

  While the LIA standard provides a set of proposed bindings for
  \CL{}, the format and depth of the specification documents is not
  readily usable by a \CL{} programmer, should an implementation
  decide to comply with the provisions.  Moreover, much latitude is
  left to each implementation to provide the LIA ``environmental''
  setup.

  It would be most beneficial if more precision were agreed upon by
  the \CL{} community about how to provide LIA compliance in the
  implementations.  In that case, a new set of documentation or
  manuals in the style of the HyperSpec could be provided, for the
  benefit of the \CL{} programmer.

  The goal of this paper is to foster a discussion within the \CL{}
  community to converge on a complete specification for LIA
  compliance.  The paper discusses some of the issues that must be
  resolved to reach that goal, e.g., error handling and full
  specification of mathematical functions behavior. 
  
\end{abstract}

\maketitle




\section{Introduction}

An interesting exercise (academic or not) that a programmer (\CL{} or
not) may find intriguing is to implement an \emph{Interval Arithmetic}
(IA) library\footnote{See, for example \cite{hickey:interval:2001}, or
  \cite{kulisch:complete:2009} for a rather complete summary with
  references to the seminal works in the area. IEEE has also published
  a preliminary standard for IA \cite{revol:introIEEEIA:2017}.}.
Programmers of all stripes would learn a lot if they tried to really
implement an IA library. But since most probably won't, this paper may
serve as sufficient summary to get you through a cocktail party
conversation on the matter.

The usefulness of IA is rather established; many numerical issues can
be naturally dealt with by using an IA library, albeit at a slight
increase in computation times. There is a nice body of literature and
proposed standards to ensure availability of IA in a computing
environment, and many of these eventually provide one or two different
\emph{interval} representations (\emph{endpoint} and \emph{midpoint}),
the operations on them, and nitpicking treatment of corner cases;
e.g., intervals with infinite endpoints and interval division by an
interval containing 0.

As we shall see, the ``nitpicking'' boils down to using the
\IEEEFPStd{}/\IECFPStd{} standards \cite{IEEE-754:2008}.  Eventually,
in this work, the use of the IEC \emph{Language Independent
  Arithmetic} (LIA) standards \cite{LIA1:2012,LIA2:2001,LIA3:2004}
will be advocated.  The LIA standard is a comprehensive collection of
concepts and carefully thought out behaviors a basic library of
\emph{integer}, \emph{floating point} and \emph{complex} numbers
mathematical functions and ancillary environment functionalities
should look like.  One of the, in the opinion of the writer, unstated
goals of LIA is that a programmer should be able to relatively easily
understand mathematical software writeen in any language ecosystem
that abided the specification.

A word of caution.  The present paper is neither a full blown
\CL{} LIA specification, nor a description of an implementation of the
functionalities depicted herein.  It rather is a \emph{leaflet} that
intend to present the community a project which, in the modest opinion
of the writer, should be completed after careful debate and careful
consideration of all the details.

\subsection{An IA Library\ldots Hitting the Wall}

An IA library in \CL{} implementing what is known as an \emph{endpoint
  representation} can be easily started as follows.  For brevity,
since it is a valid \CL{} identifier, we use the name \code{[]} here for
what other languages might call an \code{interval}.

\begin{alltt}
  (defstruct ([] (:constructor [] (low high)))
    (low 0.0 :type real)
    (high 0.0 :type real))

  (defun radius (i) (- ([]-high i) ([]-low i)))

  (defun pointp (i) (= ([]-high i) ([]-low i)))

  (defmethod add ((i1 []) (i2 []))
    ([] (+ ([]-low i1) ([]-low i2))
        (+ ([]-high i1) ([]-high i2))))

  (defmethod sub ((i1 []) (i2 []))
    ([] (- ([]-low i1) ([]-high i2))
        (- ([]-high i1) ([]-low i2))))
\end{alltt}

After starting in earnest, a \CL{} (or \Java{}, \CLang{}, \RLang{},
\Python{}) programmer is soon faced with a number of numerical issues,
should she be willing to achieve the best possible behavior out
of the IA library.

The problem is that, as mentioned before, IA specifications are usually
formulated in terms of the \IEEEFPStd{} standard, which, at this point
is readily available only to \CLang{} and \Cpp{}
programmers\footnote{There are, e.g., \Python{} bindings to
  \IEEEFPStd{}, but they rely on the underlying \CLang{} library
  implementation.\\Cfr., \texttt{https://www.python.org/dev/peps/pep-0754/}.}  In particular, the IA
specifications exploit \emph{rounding modes} and \emph{infinities},
which are unevenly available in \CL{} implementations; another,
related issue is the treatment of \emph{floating point exceptions}.


\paragraph{Rounding Modes.} If we had available some ways of handling
infinities and rounding modes, we could write the IA library
operations as follows:

\begin{alltt}
  (defmethod add ((i1 []) (i2 []))
    ([] (\codelia{rounding-down} (+ ...))
        (\codelia{rounding-up} (+ ...))))
\end{alltt}

\noindent
where \code{rounding-down} and \code{rounding-up} are macros with an
intuitive semantics.  Unfortunately, at this point in time, it is not
possible to provide the \code{rounding-down} and \code{rounding-up}
macros without delving deeply in an implementation.

\paragraph{Infinities and NaNs.} Another issue is the handling of
\emph{special values}: essentially \emph{infinities} and \emph{NaN}s
(not-a-number).  Both items are handled unevenly in \CL{}
implementations; infinities and \emph{quiet} NaNs (\cfr{} the
\IEEEFPStd{} standard) are somewhat supported; \emph{signaling} NaNs
not so much so.

\paragraph{Handling of Floating Point Exceptions.} Apart from our
doomed IA library, another issue that is not always very well
clarified in \CL{} implementations (and especially \emph{across} them)
is how floating points exceptions are handled.  The \CL{} standard
defines the conditions:

\begin{alltt}
  {floating-point-overflow}
  {floating-point-underflow}
  {floating-point-inexact}
  {floating-point-invalid-operation}
  {division-by-zero}
\end{alltt}

\noindent
Alas, their use is inconsistent across implementations (apart from the
mostly clear cut case of \code{division-by-zero}).  Two
implementations may choose to signal\\
\code{floating-point-invalid-operation} or\\
\code{floating-point-inexact} on the same operation.

This is not the only issue vis-a-vis \CL{} and the \IEEEFPStd{}.  A
deeper issue pertains the \emph{notification} machinery that is
invoked when one of the aforementioned conditions is to be signaled
by an operation. Should an implementation actually \emph{signal} a
(floating point) condition using \code{error}, or should it go the
\CLang{} way \cite{C18:2018} and \emph{record} somewhere an
\emph{indication} that a condition ``happened'', for the programmer to
check directly?

\subsection{\CL{} Implementations and the \code{:ieee-floating-point}
  Feature}

The ANSI \CL{} Standard \cite{ANSICL:1994} makes provisions for a
\CL{} implementation to ``declare'' that it \emph{purports to
  conform} to the requirements of the \emph{IEEE Standard for Binary
  Floating-Point Arithmetic} (no reference given).
There are a few problems with this statement\footnote{A form
  of ``left to the implementation'', which, as usual, does not bode
  well for the programmer.}.

The presence of the \code{:ieee-floating-point} feature in a \CL{}
implementation is a (very) partial indication that \emph{some} support
for the \IEEEFPStd{} is available.  Table \ref{tab:cl-ieee-support}
shows a summary of the current state of compliance for a number of
implementations\footnote{The table is incomplete because not all
  implementations were checked and because the notion of
  ``compliance'' is rather complicated to assess in this case.} with
respect to the notions just described.

\begin{table*}[ht!]
  \centering
  \begin{tabular}{|l|c|c|c|c|c|c|}\hline
    & CMUCL/SBCL & LW & ACL & ABCL & ECL & CCL\\\hline\hline
    Infinities      & Y & Y & Y & U & U & Y\\\hline
    Quiet NaNs      & Y & Y & Y & U & U & Y\\\hline
    Signaling NaNs  & Y & N & N & U & U & U\\\hline
    Rounding        & Y & N & N & U & U & N\\\hline
    Exceptions NACF & P & P & P & P & P & P\\\hline
    Exceptions NRI  & Y & N & N & U & U & N\\\hline
  \end{tabular}
  \caption{\small \CL{} implementations ``compliance'' status
    w.r.t. the \IEEEFPStd{}. The acronym NACF stands for
    \emph{Notification by Alteration of Control Flow}, while the
    acronym NRI stands for \emph{Notification by Recording of
      Indicators} (cfr., \cite{LIA1:2012,LIA2:2001,LIA3:2004} ); they
    will be discussed later on.  The entries are Yes, No, Unknown, and
    Partial.}
  \label{tab:cl-ieee-support}
\end{table*}

\paragraph{Infinities and NaNs.} Many implementations provide
infinities and NaNs, but with obviously different lineages.  E.g., the
following syntax is used by LW and CCL, which is then declined in
various interesting ways:\\[2mm]
\begin{tabular}{ll}
  $\infty$ & \code{1F++0}, \code{1D-+0}\\
  NaN &\code{1F+-0}, \code{1S--0}
\end{tabular}

\vspace{2mm}

\noindent
but ACL chooses to provide ``variables'' with read-time syntax.
\begin{alltt}
  ACL prompt> \codeprompt{*infinity-single*}
  \textit{\#.*INFINITY-SINGLE*}

  ACL prompt> \codeprompt{(+ 42 *nan-single*)}
  \textit{\#.*NAN-SINGLE*}
\end{alltt}

\noindent
While this may be perfectly sensible it has the drawback of not
playing so nicely with \code{*read-eval*}.

The only implementations that allow a programmer to get a handle on a
\emph{signaling} NaN are CMUCL and SBCL.  There appear to be no easy
way to create such a value in the other implementations.

\paragraph{Rounding.} Only CMUCL and SBCL allow the setting of the
rounding mode by accessing directly the equivalent of the \IEEEFPStd{}
\emph{floating point environment}.  However, the facility -- which
resembles the \CLang{} library \code{fenv.h} setup -- is very low
level.


\subsection{A Proposal}

Alas, ``just adding'' infinities and rounding modes to a \CL{}
implementation may not not be quite sufficient, as their semantics is
deeply intertwined with the other parts of the \IEEEFPStd{} standard.
A better course of action would be to nudge the implementors to comply
with the current standards. The definition of a new \CL{}
``arithmetic'' specification may be a better way to achieve the goal
of providing \CL{} programmers with a layered set of documented
functionalities.

The observation is that by now, the \IEEEFPStd{} standards (and to a lesser
extent the LIA standards) appear to be quite accepted and common place
in many programming language ecosystems. It is the opinion of rhe
writer that for the \CL{} community, ``to go with the flow'' would be
the pragmatic thing to do.




\section{Goals and Issues}

The goal of this paper is to urge the various \CL{} implementations to
provide better support for floating point (hence complex) arithmetic,
in order to make it possible to directly write a IA library (and other
numerical routines) in an easier and and more \emph{correct} way.
The point of view is that of a \CL{} programmer and user. The main
source of this proposal are the \emph{Language Independent
  Arithmetic} (LIA) specifications
\cite{LIA1:2012,LIA2:2001,LIA3:2004}, which incorporate
\IEEEFPStd{}/\IECFPStd{} \cite{IEEE-754:2008}.

\subsection{The LIA Specifications and \CL{}}

The LIA specifications are three documents covering the following
topics.
\begin{description}
\item[LIA Part 1:] integer and floating point arithmetic
\item[LIA Part 2:] elementary numerical functions
\item[LIA Part 3:] complex integer and floating point arithmetic
\end{description}
The LIA specifications take great care not to be overly constraining
(they relax a few requirements of \IEEEFPStd{}/\IEEEFPStd{}) while
being very precise about the behavior of each item they define.  They
also contain appendices describing suggested bindings for various
languages, \CLang{}/\Cpp{} and \Fortran{} being prominent, including
\CL{}.

A \CL{} implementor could, in principle, just read through the LIA
specifications and provide all the necessary bits and pieces while
building the arithmetic facilities of the language. Yet, it is the
writer's opinion that this course of action would still fall a bit
short of providing a programmers' computing environment with an
``implementation independent'' firm ground.  The reasons lie in the
LIA specifications themselves, as they understandably cannot provide
more than a suggestion about how a language binding should cover and
look like.  There are some issues that a more \CL{} centric
specification would and should clarify: \emph{naming conventions},
\emph{layering and packaging}, \emph{programming environment setup},
\emph{rounding-modes} and, above all, \emph{error handling} and the
\emph{floating point environment}.  In the following each of these
issues will be discussed. Eventually, the result should be an in-depth
specification formatted in the style of the \CL{} standard
\cite{ANSICL:1994,ANSIHyperSpec:1996}.

\subsubsection{Naming Conventions}

The LIA specifications suggest a naming convention for its
functionalities that reuses much of \CL{} names.  some of the choices
are not particularly in line with \CL{} style.  Two examples are the
functions \code{sqrtUp} and \code{sqrtDwn}, which compute square roots
with ``up'' or ``down'' rounding modes; \CL{} style would have avoided
the ``camel case'', given that \CL{} implementations are uppercasing
out-of-the box, while preferring an hyphenated naming.

Another issue with the LIA suggested naming is that it essentially
requires an implementation to provide a set of very basic
LIA-compliant functions -- e.g., \code{+}, \code{*}, \code{1-},
\code{sin}, etc. -- which implies a reworking of an implementation
core.

\subsubsection{Layering and Packaging}

In order to provide a \CL{} centric LIA specification and adoption
path, it would be better to ensure that the new functionality were
provided as a library.  This means, at a minimum, to provide a package
that contained all the ``new'' names introduced.  Given the partition
of the LIA specifications, further sub-packaging could be provided.

A first cut proposal would be to have a package named (or nicknamed)
\code{CL-LIA} that exported all the names that are necessary to
implement a form of the LIA specifications.  As it will be discussed
later on, it will be useful to have a
\code{cl-lia:floating-point-invalid-operation} condition, despite the
presence of the standard \CL{} one.

\subsubsection{Programming Environment Setup}

The \IEEEFPStd{}/\IECFPStd{} and LIA specifications define a number of
environment checks that a compiler or a program may check to produce
code that complies and/or exploits their semantics.  These are akin to
the \code{:ieee-floating-point} feature (which, as we have seen, is
only partially informative).   At least two sets of
``checks'', both in functional and ``feature'' form could be provided
by a \CL{} LIA implementation.

\paragraph{Library Compliance Checks.} The LIA specifications require
a boolean variable \code{iec60559}$_F$ that reports whether or not an
implementation complies with the \IEEEFPStd{}/\IECFPStd{}
implementation of the floating point type $F$.  This is more stringent
than the ``bulk'' statement implied by the \code{:ieee-floating-point}
feature, although LIA1 (cfr, \cite{LIA1:2012} Section 5.2) explicitly
states that no exact floating point representation is required.  A
\CL{} LIA implementation should define similar constants, boolean
functions and, possibly, features.

\paragraph{Layered Library Checks.} Another set of variables, boolean
functions and features should be made available to indicate the level
of compliance with the LIA specifications.  A suggestion is to provide
the functions (and therefore constants and features)
\code{LIA1-compliance}, \code{LIA2-compliance}, and
\code{LIA3-compliance}.  Finer statements may pertain parts of each
specification; one example is the \code{provides-infinities-p} and
\code{provides-nans-p}.  Other such checks can be described for other
parts of the \CL{} LIA implementation, as seen below for
\emph{exception handling}.  Finally, one important check that could be
provided is whether the \CL{} implementation carries over the LIA
semantics to the functions in the \code{COMMON-LISP} package: the
check \code{is-cl-using-lia} would state that a function like, for example,
\code{cl:sin} implements the LIA semantics w.r.t. infinities and NaNs,
rounding modes and exception notifications (see below).  Of course,
whether or not provide such ``history rewriting feature'' is up for
debate.  One argument in favor is that old code would still work without
having to be tweaked to use the new functions provided in a
\code{CL-LIA} package.

\subsubsection{Rounding Modes}

Floating points numbers being approximation of real numbers carry with
them notions of \emph{rounding}.  The LIA specifications define how
rounding modes affect elementary and library operations.

Following in this case the \CLang{}/\Cpp{} example, it could be
possible to define a set of constants with the meaning showed in Table
\ref{tab:round-consts}.  The type \code{rounding-mode} can then be
defined as:


\begin{alltt}
  (deftype \codelia{rounding-mode} ()
    `(member ,indeterminate
             ,to-zero
             ,to-positive-infinity
             ,to-negative-infinity
             ,to-nearest-even))
\end{alltt}
          
\noindent
The rounding mode can then be tracked using a special variable
\code{*rounding-mode*}.  A macro \code{with-rounding-mode} is an
obvious extension as well as macros wrapping one expression:
\code{round-upward}, \code{round-downward}, \code{round-nearest} etc.

\begin{table}
  \centering
  \begin{tabular}{lrl}
    \CL{} constants & Value & LIA meaning\\\hline
    \code{indeterminate} & -1 & {\ }\\
    \code{to-zero}     &  0 & truncate\\
    \code{to-nearest}  &  1 & nearest\\
    \code{to-positive-infinity} & 2 & other\\
    \code{to-negative-infinity} & 3 & other\\
    \code{to-nearest-even}      & 4 & nearesttiestoeven\\
  \end{tabular}
  
  \caption{Proposed \CL{} constants representing rounding modes.}
  \label{tab:round-consts}
\end{table}

Moreover, the LIA specifications define some functions that guarantee
a given rounding result; e.g., there are three versions of
$\sqrt{\cdot}$, which compute square roots guaranteeing rounding
upward, downward and to nearest.  For \CL{} it will probably be better
to provide \emph{four} such ``names'' with the following, LIA-inspired,
suffixes: \code{sqrt} (no suffix), \code{sqrt.<}, \code{sqrt.<>}, and
\code{sqrt.>}. The first version depends on the current rounding mode,
the other ones round down, near and up (suffixes \code{.<}, \code{.<>}
and \code{.>}).

\subsubsection{Error Notification/Handling and the Floating Point
  ``Environment''}

The LIA specifications must address the differences (and \ldots
``traditions'') that different communities have developed over the
years.  The exegesis of the LIA specifications seems to point out that
the major concern was to disentangle concerns that earlier language
specifications (especially the \CLang{}/\Cpp{} ones) addressed in a
idiosyncratic way.  Within the \CL{} community, the Condition System -
as prefigured and hashed out in \cite{pitman:1990:exceptional} --
offers all the bells and whitles to implement the programmer's side of
error handling, but some issues must be dealt with at the
implementation level.

One of the main tangled issues regards the \emph{handling of
  ``errors''}, that is, after what an ``error'' is agreed upon. This
is a notoriously complicated issue which the LIA specifications appear
to break down into two parts.
\begin{itemize}
\item How errors are \emph{notified}.
\item What happens depending on the \emph{notification style}.
\end{itemize}
The LIA specifications assume that a language ``environment''
establishes some forms of \emph{notification} machinery.  Three major
modalities are singled out.
\begin{enumerate}
\item \emph{Notification by recording in indicators} (\textbf{NRI} --
  LIA1, Section~6.2.1).
\item \emph{Notification by alteration of control flow} (\textbf{NACF}
  -- LIA1, Section~6.2.2).
\item \emph{Notification by termination with message} (\textbf{NTM} --
  LIA1, Section~6.2.3).
\end{enumerate}
The LIA1 specification, Annex~D, proposes that \CL{} defined the
arithmetic exception handling using the Condition System, i.e., using
the NACF notification approach.  However, \SBCL{}/\CMUCL{} provide --
de facto -- a NRI setup modeled on the \CLang{} NRI interface
provided by \code{<fenv.h>}.  It would be better to accommodate
\textbf{both} alternatives NRI and NACF, and make them available to
the programmer for finer control.

LIA1 provides an example about how \Fortran{} may provide some
compiler directives to choose between NRI and NTM (cfr., LIA1, Annex
E).
\begin{alltt}
  !LIA\$ NOTIFICATION=RECORDING
  !LIA\$ NOTIFICATION=TERMINATE
\end{alltt}

In order to select and introspect what kind of exception handling
regime is in place in a given computation\footnote{The LIA
  specifications make no mention of any \emph{threading model};
  however, it is assumed that an implementation can make all the
  dynamical behavior of numerical computations ``thread safe''.} an
appropriate \CL{} API will have to be defined.

\noindent
To complete the discussion, we must consider the \emph{floating point
  environment}, \emph{conditions and continuation values}, and
\emph{underflow/overflow}.

\paragraph{Floating Point Environment.} Having control over the kind
of notification style is nice, but it requires a better handling of
the \emph{floating point environment}, which \CLang{} handles through
\code{<fenv.h>}, and that \CMUCL{}/\SBCL{} manipulate using a few
functions and what looks like is an a-list.

The floating point environment is used as a kitchen sink to keep track
of rounding modes, exceptional situation notifications and other
information.  A unified item representing these concerns still seems
the best way to give access to them in a dynamic way.

\paragraph{Conditions and ``Continuation'' Values.} The operations
that operate on the ``borderline'' case in LIA (e.g., operations on
\emph{NaN}s, or that generate \emph{underflow}s and \emph{overflow}s,
are specified alongside a \emph{continuation value}.  This is most
important for the NRI notification style, where an operation
``continues'', while recording the indication of an exceptional
situation.  To facilitate the implementation of a LIA-compliant
package for \CL{}, it would be useful to mirror the
\code{arithmetic-error} sub-hierarchy and to equip the classes with a
\code{continuation-value} slot (alongside appropriate initargs and
readers).

\section{Considerations for a New \CL{} Arithmetic Specification}

Having discussed some of the issues about providing support for the
LIA specification in \CL{}, we here offer a detailed opening bid in a
hoped-for public discussion on the creation of a \CL{} binding, or
otherwise integrating its ideas into the language.

A full-blown document containing the full description of each item, in
a style reminiscent of the \CL[] ANSI Standard \cite{ANSICL:1994}, is
in the works.  The LIA specification describes each item and,
especially, operation, in a terse and abstract way, which requires
quite a bit of effort to map onto a typical \CL{} description,
especially for functions results.  The full blown description is
intended to be read by a \CL{} programmer.

\paragraph{Package.}
All the names that will be introduced or shadowed
from the \code{"COMMON-LISP"} package will be exported from a new
package.  The proposed nickname is \code{"CL-LIA-MATH"}\footnote{Or
  \code{"CL.MATH"}, should a more ambitious naming be adopted.}.

\paragraph{Environmental Features and Switches.}
An implementation will state what parts of the specification will be
available. The following (semi-hierarchical) set of environmental
queries will be available as \emph{functions}, \emph{special
  variables}, and/or \emph{features}.
\begin{alltt}
  lia-subset-available
    lia1-subset-available
    lia2-subset-available
    lia3-subset-available
  lia-compliance
    lia1-compliance
      provides-infinities
      provides-nans
      provides-rounding-modes
      provides-floating-point-environment
      provides-nacf
      provides-nri
      provides-ntm
    lia2-compliance
      cl-package-uses-lia
    lia3-compliance
\end{alltt}
The above list represents Boolean functions.  The
\code{provides-\ldots} functions return \textit{true} when the specific
functionality is \emph{fully provided}. In the above example
\code{lia1-compliance} returns \textit{true} when \emph{all} the
\code{provides-\ldots} functions return \textit{true}.

Note that, while the list of introspective facilities listed covers
most of the dimensions in LIA compliance, certain combinations are
ruled out by the detailed specification and by the way it is presented
in the standards.  E.g., it is not very sensible to have an
implementation for which\\
\code{provides-floating-point-environment} returned \textit{false},
while \code{provides-nacf} returned \textit{true}.

\paragraph{Infinities, NaNs and Rounding.}
An implementation of the specification will offer all the values
pertaining \emph{infinities}, \emph{NaNs} and \emph{rounding}.  In
particular, a fully LIA compliant will provide the environment
introspection functions and variable mandated by LIA1 and ``typed''
constants like\\\code{double-float-positive-infinity} and
\code{infD}.  Moreover, the full specification will clarify the
behavior of each function and operation when presented with
\emph{NaNs}, both \emph{quiet} and \emph{signaling}, especially
regarding the interplay with the \emph{error notification style} (see
below).

The rounding versions of all the LIA mandated operations will be
marked by the postfixes \code{.<}, \code{.<>}, and \code{.>},
signifying rounding towards negative infinity, nearest, and positive
infinity.  The macro \code{rounding} has effect on the ``unqualified''
versions of the arithmetic operations.  E.g.\\[2mm]
\begin{tabular}{lcll}
  \code{(\codeprompt{+.<} pi pi)}
  & $\Rightarrow$
  & $2\pi$
  & rounded toward $-\infty$.\\
  \code{(\codeprompt{+.>} pi pi)}
  & $\Rightarrow$
  & $2\pi$
  & rounded toward $+\infty$.\\
\end{tabular}

\vspace*{2mm}

\noindent
\ldots, while using the \code{rounding} macro\\[2mm]
\begin{tabular}{l}
  \code{(\codeprompt{rounding :positive-infinity}
  (\codeprompt{+} pi pi))}\\
  $\Rightarrow$ $2\pi$ rounded toward $+\infty$.\\
\end{tabular}

\vspace*{2mm}

\noindent
\ldots, but\\[2mm]
\begin{tabular}{l}
  \code{(\codeprompt{rounding :positive-infinity}
  (\codeprompt{+.<} pi pi))}\\
  $\Rightarrow$ $2\pi$ rounded toward $-\infty$.\\
\end{tabular}

\vspace*{2mm}

\noindent
I.e., the \code{rounding} macro establishes a dynamic environment with
a specific rounding mode set up, which can be ignored by the ``hard
rounding'' versions of the operations in the body.

\paragraph{Floating Point Environment and Error Handling.}
An implementation of the specification shall assume the presence of an
\emph{opaque} data type called \code{floating-point-environment}.  The
access functions for this data structure are patterned after the
\CLang{}/\Cpp{} standards in order to offer familiarity and, possibly,
ease of implementation.

An implementation of the specification will always offer \emph{both}
NACF and NRI notification styles, with full control offered to the
programmer about when and where they turn on and off each style.  The
NTM notification style will used only for catastrophic events, which
will be documented accordingly.

The type \code{arithmetic-notification-style} can be defined as:

\begin{alltt}
  (deftype \codelia{arithmetic-notification-style} ()
    '(member :recording   \textcolor{red}{; I.e., NRI.}
             :error       \textcolor{red}{; I.e., NACF.}
             :terminating \textcolor{red}{; I.e., NTM.}
             ))
\end{alltt}

The main functions and macros that allow full control of the
notification style and \emph{continuation values} possibly returned by
an operation are the following.
\begin{alltt}
  current-notification-style
  set-notification-style
  with-notification-style
  trap-math
\end{alltt}

\vspace*{2mm}

\noindent
The \code{trap-math} macro is
intended as a wrapper around the \CL{} error handling machinery
(\code{handler-case}, \code{unwind-protect}, etc\ldots) that
automated some of the setup and teardown operations on the floating
point environment, alongside the handling of continuation values. A
possible syntax is the following
\begin{alltt}
  (trap-math (<\textit{options}>)
      <\textit{expr}>
      <\textit{handler}>* )
\end{alltt}
The \codeit{options} parameter is a list that may contain the keywords
\code{:notify-by},  \code{:before}, and \code{:after}.
The \code{:notify-by} is the \emph{notification style} defaulting to
\code{:error}, \code{:before} and \code{:after} instead demarcate
lists of \emph{actions} that intend to simplify the setting up and the
teardown of indicators in the floating point environment: \code{:save}
saves the current floating point environment, \code{:clear} creates a
fresh floating point environment with no indicators recorded, and
\code{:merge} (in an \code{:after} position) merges the current
floating point environment with the possibly saved one. Of course, a
different syntax is possible, and it is unclear to the writer which
would be the best; consider the following alternative.
\begin{alltt}
  (trap-math (\&key \textit{notify-by} \textit{before} \textit{after})
      <\textit{expr}>
      <\textit{handler}>* )
\end{alltt}

\noindent
The \codeit{handler} is a simplified list that has the following
syntax.
\begin{alltt}
  (<\textit{aec}> (\&optional <\textit{varname}>)
       \&rest <\textit{actions}>)
\end{alltt}
where \codeit{aec} is an \code{arithmetic-condition} carrying a
continuation value, \codeit{varname} is a symbol that can be bound to
the condition instance and \codeit{actions} is a list that may contain
the following items.
\begin{itemize}
\item \code{:default} -- the behavior is the standard one for
  \codeit{aec}.
\item \code{:clear} -- when combined with the \code{:continue} forms
  and the complex \code{:error} form,
  it clears the indicator corresponding to \code{aec}
  from the floating point environment.
\item \code{:raise} -- re-signals the \code{aec}
\item \code{(:raise \codeit{c} \&rest \codeit{args})} -- signals a
  new contition \codeit{c}.
\item \code{:continue}, \code{(:continue \textit{expr})} --
  continues the computation by yielding the standard continuation
  value of the result of evaluating \codeit{expr}; the
  \code{:continue} actions can be rendered by means of
  \code{cl:use-value} and/or \code{cl:continue} restarts.
\end{itemize}
An example of the use of \code{trap-math} is the following, which is
also a rendering of \cite{LIA1:2012} Appendix A.6.
\begin{alltt}
  (\codelia{trap-math} (\codelia{:before :save :clear}
              \codelia{:after :merge})
      (fast-solution \textit{input})
      (cl:floating-point-overflow ()
         \codelia{:clear}
         (\codelia{:continue} (reliable-solution input))))
\end{alltt}
or, with a different syntax
\begin{alltt}
  (\codelia{trap-math} (\codelia{:before} (\codelia{:save :clear})
              \codelia{:after :merge})
      (fast-solution \textit{input})
      (cl:floating-point-overflow ()
         \codelia{:clear}
         (\codelia{:continue} (reliable-solution input))))
\end{alltt}

\subsection{Providing the Specification -- A Descriptive Example}

\begin{figure*}[ht!]
  
  \begin{minipage}{.8\textwidth}
\DDictionaryItem{Function \code{=}, \code{/=}}
\index{*!\code{=}}
\index{*!\code{/=}}

\DSyntax{}

\code{=} \varname{a}, \varname{b} \RArrow \varname{boolean}\\
\code{=} \varname{a} \code{\&rest} \varname{bs} \RArrow \varname{boolean}\\
\code{/=} \varname{a}, \varname{b} \RArrow \varname{boolean}\\
\code{/=} \varname{a} \code{\&rest} \varname{bs} \RArrow \varname{boolean}

\DArgsNValues{}

\varname{a} \varname{b} -- Numbers.\\
\varname{bs} -- A list of numbers.\\
\varname{boolean} -- a \clterm{generalized boolean}.

\DDescription{}

The dyadic version of \code{=} (and \code{/=}) performs an arithmetic
equality (inequality) test between \varname{a} and \varname{b}.  The
monadic and n-adic versions are built upon the dyadic one as per the
regular \CL{} description in \cite{ANSIHyperSpec:1996}.

It is assumed that \varname{a} and \varname{b} are converted (as per
the \emph{contagion rules} of \CL{}) to be of the same type.
Therefore the following cases can be be considered as per the LIA
specifications.

\begin{enumerate}[label=(\alph*)]
\item If \varname{a} and \varname{b} are either finite integers, finite
floating point numbers, or finite complex numbers then the result is
\varname{true} (respectively, \varname{false}) if the two numbers are
equal (respectively, different) in the mathematical sense.  In the
LIA spec this is the result of $\mathit{eq}_T(a, b) \equiv a = b$ or
$\mathit{neq}_T(a, b) \equiv a \neq b$ for an
appropriate $T$.  This is the standard \CL{} case.

\item If \varname {a} and \varname {b} are \clieeeterm{infinities} then
\code{=} returns \varname{true} (respectively \varname{false}) if they
are both positive or both negative; otherwise it returns
\varname{false} (respectively \varname{true}).

\item If either \varname {a} or \varname {b} is a \clieeeterm{quiet NaN},
and, respectively, \varname {b} and \varname {a} is not a
\clieeeterm{signaling NaN}, then the result is \varname{false}.

\item Complex numbers are checked recursively on the real and imaginary
  parts.
\end{enumerate}

\DExceptional{}

If either \varname {a} or \varname {b} is a \clieeeterm{signaling
  NaN}, then, under the notification NACF regime, the indicator
\code{:invalid} is recorded and the
\code{floating-point-invalid-operation} is signaled (with
\emph{continuation value} \code{NIL} recorded); otherwise, under the
NRI notification regime, the indicator \code{invalid} is recorded and
\code{NIL} (\varname{false}) is returned as \emph{continuation value}.

For complex numbers, the recording and signaling operations (under NRI
and NACF) happens if the condition above applied to either of the real
or the imaginary parts of \varname{a} and \varname{b}.\\[1mm]
\hrule
\end{minipage}
\caption{An example entry that should appear in the full specification
  for the \CL{} LIA-compliance documentation.}
\label{fig:example-spec}
\end{figure*}


Eventually, the considerations put forth in this paper should be
crystallized in a specification that clarified all the many thorny
issues that will crop up when considering as many details as possible.
The goal will be to provide a specification \emph{\'{a} la} \CL{}
HyperSpec \cite{ANSIHyperSpec:1996}, where each item (function,
variable, class, etc.) has a mostly self-contained description with,
by now, a conventional structure.  This is different from the
presentation style adopted by the LIA specifications, which
heavily rely on quite formal yet ``generic'' description of each
operation's behavior.

The most important aspects a \CL{} LIA specification will be to
describe, for each function (or other item), the following behaviors.
\begin{enumerate}[label=(\alph*)]
\item The \emph{corner cases}: infinities and NaNs.
\item The interplay between the \emph{notification style}, the
  handling of errors, the floating point environment and the
  \emph{continuation values} that are specified according to the LIA
  documents.
\end{enumerate}

As an example of what an entry in the envisioned full \CL{} LIA
specification would look like, Figure \ref{fig:example-spec} shows a
description of the (dyadic) \code{=} and \code{/=}
functions. Hopefully, all details and corner cases listed above have
been taken into consideration.  The reader can compare the \emph{equality}
specification in the LIA1 document with the one in Figure
\ref{fig:example-spec}.

\section{Conclusions and Final Discalimers}

In order to write a fully functional (according to the available
literature and proposed standards) IA library in \CL{}, a programmer
needs a finer control over the \emph{floating point} environment and
access to functionalities such as rounding modes.

This paper puts forth a proposal to complete a \CL{} HyperSpec styled,
LIA-based specification that would provide a more accessible
documentation for a programmer and a clear guideline about how certain
functionalities should be provided by an implementation.

Given an implementation of the proposed ``New Arithmetic
Specification'' a programmer could at least start to write a proper IA
library.  As an example, the \code{add} method would look like the
following.
\begin{alltt}
  (defmethod add ((i1 []) (i2 []))
    ([] (\codelia{+.<} ([]-low i1) ([]-low i2))
        (\codelia{+.>} ([]-high i1) ([]-high i2))))
\end{alltt}
where \code{+.<} and \code{+.>} are the addition operations on floating
point numbers that round, respectively, \emph{downward} and
\emph{upward}.

Again, the writer wants to insist and repeat that the present paper is
a \emph{leaflet} that intends to present the \CL{} community a project
which, in his modest opinion, should be completed after careful debate
and careful consideration of all the details within the \CL{} community.

A full blown specification covering LIA-1, LIA-2 and LIA-3 will run
close to two hundred pages if written and formatted according to the
\cite{ANSIHyperSpec:1996} style (a worthy goal in itself).  Many of
the examples contained in this paper are \emph{suggestions} about how
they could look. Agreement withing the \CL{} community will help
settle down several issues this paper puts forth.

What this paper wants to point out though, is that many researchers
and practitioners did lay down a sensible set of standards, the LIA
standards, which \emph{did} take into account \CL{}.  Following them
appears to be one good way to ensure that \CL{} will keep its place
among the most important language ecosystems around, and welcome
programmers from other communities by offering them a familiar playpen
and more.


\bibliographystyle{plain}
\bibliography{ELS-IA-LIA}

\end{document}